Technical University Dresden

MASTER THESIS

# The Marriage of Incremental and Approximate Computing

**Author:** Dhanya R Krishnan

**Course:** MSc. Distributed Systems Engineering

**Supervisor:** Dr. Pramod Bhatotia

**Professor:** Prof. Dr. Christof Fetzer

**Submitted:** 12$^{th}$ March 2016

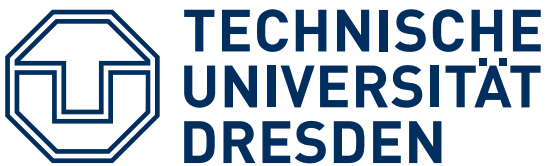

# Declaration

I declare in lieu of an oath that the Master Thesis submitted has been produced by me without illegal help from other persons. I state that all passages which have been taken out of publications of all means or un-published material either whole or in part, in words or ideas, have been marked as quotations in the relevant passage. I also confirm that the quotes included show the extent of the original quotes and are marked as such. I know that a false declaration will have legal consequences.

**Date:** ___________________. **Signature:** _______________________________.

# Acknowledgment

I would like to thank my advisor Dr. Pramod Bhatotia for his guidance and constructive criticism that has helped me understand the research field better. He has been a true mentor and his high standards of quality were really inspiring and continuously motivative. I have learned many things from him including the basics of framing a research idea to tackling the disruptive research questions that come along. I would also like to thank him for his professional advice from the depth of his experiences.

I would like to thank Do Le Quoc who has been my research partner and friend. I appreciate him for his insightful thoughts and comments which has made this thesis an enjoyable work for me. I would also like to thank Prof. Christof Fetzer and Prof. Rodrigo Rodrigues for their valuable comments. Prof. Rodrigo's detailed feedbacks and comments have helped me improve the work further. I have high regards for Dr. Wiltrud Kuhlisch who was really kind to discuss the correctness of the statistics work in this thesis. I highly appreciate her for offering help regardless of our language barriers.

I have made some great friends in Germany and I am thankful to them for the great time I had outside work. Sankalp Dhasmana has been a great friend and guide, encouraging me to be resilient always. I would also like to thank all my other friends for their help and support.

Above all, my sincere gratitude to my parents who have been my tremendous source of motivation and have always let me chase my dreams. Mom and Dad, you are my treasure.

# Abstract

Most data analytics systems that require low-latency execution and efficient utilization of computing resources, increasingly adopt two computational paradigms, namely, incremental and approximate computing. Incremental computation updates the output *incrementally* instead of re-computing everything from scratch for successive runs of a job with input changes. Approximate computation returns an *approximate* output for a job instead of the exact output.

Both paradigms rely on computing over a subset of data items instead of computing over the entire dataset, but they differ in their means for skipping parts of the computation. Incremental computing relies on the *memoization* of intermediate results of sub-computations, and reusing these memoized results across jobs for sub-computations that are unaffected by the changed input. Approximate computing relies on representative *sampling* of the entire dataset to compute over a subset of data items.

In this thesis, we make the observation that these two computing paradigms are complementary, and can be married together! The high level idea is to: *design a sampling algorithm that biases the sample selection to the memoized data items from previous runs*. To concretize this idea, we designed an online stratified sampling algorithm that uses self-adjusting computation to produce an incrementally updated approximate output with bounded error. We implemented our algorithm in a data analytics system called IncAppox based on Apache Spark Streaming. Our evaluation of the system shows that IncApprox achieves the benefits of both incremental and approximate computing.

# Contents

# List of Figures

# List of Algorithms

# 1 Introduction

Modern online services produce huge quantity of real-time data from disperse events. This real-time streaming data arrive as continuous, rapid, transient and unbounded streams. Such data is most valued at the time of arrival, and in fact, the value of this streaming data decreases with time. For example, a social network may wish to detect trending conversation topics in minutes; and a service operator may wish to monitor program logs to detect failures in seconds. For such low-latency applications, there is a requirement for real-time stream computations.

In contrast to database model where data can be stored, indexed and then processed, the streaming computation model requires analysis of data *on the fly* while it streams through the server. Furthermore, processing such huge quantity of unbounded data streams need massively parallel and distributed computing frameworks [2, 7, 21, 23, 48, 62, 75].

In order to support today's big-data driven technologies that uses streaming data, there are two desirable, yet contradictory requirements [62, 75] *(i)* achieving low latency; and *(ii)* efficient resource utilization. For instance, in order to achieve low-latency requirements, we can increase the computing resources and parallelize the application logic using a distributed infrastructure. Since most data analytics frameworks are based on the data-parallel programming model [34], almost linear scalability can be achieved with increased computing resources. However, processing all data items of the input stream would require more than the available computing resources to meet the desired SLAs or the latency guarantees.

To counterbalance these two contradictory goals, there is a surge of new computing paradigms that prefer to compute over a subset of data items instead of the entire data stream. Since computing over a subset of the input requires less time and resources, these computing paradigms can achieve bounded latency and efficient resource utilization. In particular, two such paradigms are incremental and approximate computing.

## 1.1 Incremental computing

Many data analytics jobs operate by repeatedly invoking the same application logic or algorithm over an input data that differs slightly from that of the previous invocation [24, 45, 48]. In such a workflow, small, localized changes to the input often require only small updates to the output, creating an opportunity to update the output incrementally instead of recomputing everything from scratch [9, 17]. Since the work done is often proportional to the change size rather than the total input size, incremental computation can achieve significant performance gains (low latency) and efficient utilization of computing resources [19, 22, 66].

There are two main approaches for incremental updates. First approach is to provide programmers the facility to store and use states across successive runs so that only computations which are affected by changes to the input needs to be executed [55, 64]. This strategy requires programmer to devise application-specific incremental update mechanism (or a *dynamic algorithm*) for updating the output as the input changes [25, 30, 35, 36, 40, 44]. While dynamic algorithms can be asymptotically more efficient than re-computing everything from scratch, research in the algorithms community shows that these algorithms can be difficult to design, implement and maintain even for simple problems. Furthermore, these algorithms are studied mostly in the context of the uniprocessor computing model, making them ill-suited for parallel and distributed settings which is commonly used for large-scale data analytics. Another approach is to develop systems which can re-use the results of previous computations transparently without programmer interventions. Such self-adjusting computation [9, 11, 46, 47] overcome the limitations of dynamic algorithms. At a high level, self-adjusting computation enables incremental updates by creating a dynamic dependence graph of the underlying computation, which records control and data dependencies between the sub-computations. Given a set of input changes, self-adjusting computation performs change propagation, where it reuses the *memoized* intermediate results for all sub-computations that are unaffected by the input changes, and re-computes only those parts of the computation that are transitively affected by the input change. As a result, self-adjusting computation computes only on a subset (*"delta"*) of the computation instead of re-computing everything from scratch.

## 1.2 Approximate computing

In big-data systems, due to the huge quantity of data to be processed as well as the latency and resource bounds, there is a trade-off between accuracy and speed of computation. This is based on the observation that many data analytics jobs are amenable to an approximate, rather than the exact output. Many applications like image processing, machine learning, Monte-Carlo computations etc., are compliant to approximation [37, 61, 63, 68]. For such an approximate workflow, it is possible to trade accuracy by computing over a partial subset instead of the entire input data to achieve low latency and efficient utilization of resources.

Approximate computing is a well researched field and the researchers in the database community have proposed many techniques for approximate computing. Some of the frequently used approximate computing techniques are sampling [14, 42], sketches [33], and online aggregation [49]. These techniques make different trade-offs with respect to the output quality, supported query interface, and workload. However, the early work in approximate computing was mainly targeted towards the centralized database architecture, and it was unclear whether these techniques could be extended in the context of big data analytics.

Recently, sampling based approaches have been successfully adopted for distributed data analytics [12, 43]. These systems show that it is possible to have a trade-off between the output accuracy and performance gains (also efficient resource utilization) by employing sampling-based approaches for computing over a *subset* of data items. However, these "big data" systems target batch processing workflow and cannot provide required low-latency guarantees for stream analytics.

## 1.3 The marriage

In this paper, we make the observation that the two computing paradigms, incremental and approximate computing, are complementary. Both computing paradigms rely on computing over a subset of data items instead of the entire dataset to achieve low latency and efficient cluster utilization. Therefore, we propose to combine these paradigms together in order to leverage the benefits of both. Furthermore, we achieve incremental updates without requiring the design and implementation of application-specific dynamic algorithms, and support approximate computing for stream analytics. Our work targets applications which require near real-time stream processing and are amenable to approximation.

The high-level idea is to design a sampling algorithm that biases the sample selection to the memoized data items from previous runs. We realize this idea by designing an online sampling algorithm that selects a representative subset of data items from the input data stream. Thereafter, we bias the sample to include data items for which we already have memoized results from previous runs, while preserving the proportional allocation of data items of different (strata) distributions. Next, we run the user-specified streaming query on this biased sample by making use of self-adjusting computation and provide the user an incrementally updated approximate output with error bounds.

In this thesis, we present our algorithm that combines both incremental and approximate computing with the following contributions:

- **Transparent combination of paradigms**: INCAPPROX is the first system to combine approximation and incremental computing techniques for streaming data. We use sampling and memoization techniques for achieving approximation and incremental computation respectively. Moreover, we do this transparently, decreasing the

  burden on the programmer to design and implement dynamic algorithms.

- **Query budget and Error Estimation**: We allow users to specify a query budget in terms of latency or cluster resources to be used and we guarantee this budget. Our guarantee is based on an accuracy trade-off. In order to measure the accuracy, we also provide a confidence interval or error bound on the accuracy of output.

- **Improved efficiency**: We improve the efficiency of computation by a combination of both incremental and approximate paradigms, thus achieving the benefits of both.

We implemented our algorithm in a system called INCAPPROX based on Apache Spark Streaming [6], and evaluated its effectiveness by applying INCAPPROX to various micro-benchmarks. Furthermore, we report our experience on applying INCAPPROX on two real-world case-studies: *(i)* real-time network monitoring, and *(ii)* data analytics on a Twitter stream, the details of which could be found in our paper [51]. Our evaluation of the system shows that INCAPPROX achieves a speedup of $\sim 2\times$ over the native Spark Streaming execution, and $\sim 1.4\times$ over the individual speedups of both incremental and approximate computing. This work was done jointly with Do Le Quoc and is originally published in [51].

# 2 Overview

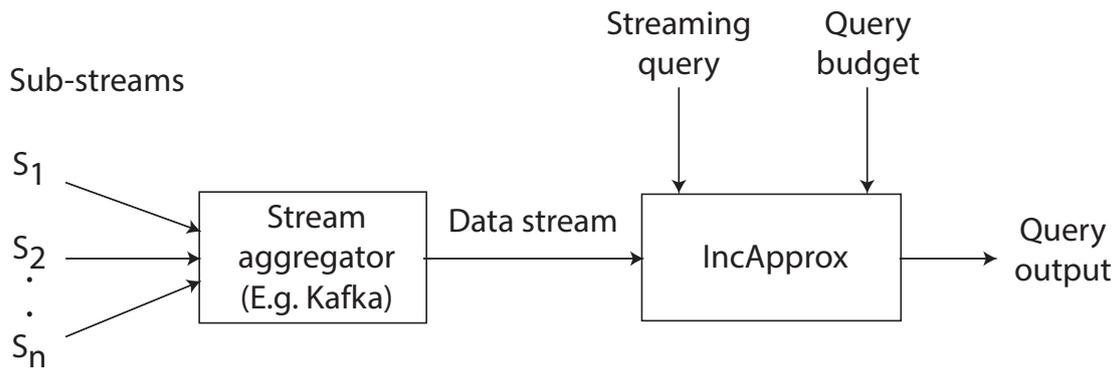

Figure 2.1: System overview

## 2.1 System Overview

The main design goal of IncApprox is real-time data analytics on online data streams. Figure 2.1 depicts the high-level design of IncApprox. The online data stream consists of data items from multiple and diverse sources of events or sub-streams. We use a stream aggregator (such as Apache Kafka [5], Apache Flume [3], Amazon Kinesis [1], etc.) that integrates data from these sub-streams. Thereafter, the system reads this integrated data stream as the input. We provide a user interface to facilitate user querying on this data stream. The user interface consists of a streaming query and a query budget. The user submits the streaming query to the system as well as specifies a query budget for processing the streaming query. The query budget can be in the form of (*i*) latency guarantees/SLAs for data processing (*ii*) desired result accuracy or (*iii*) computing resources available for query processing. Our system assures that the computation done over the data remains within the specified budget. To achieve this, the system uses a combination of incremental and approximating computing for real-time processing over the input data stream. Finally, the system emits the query result along with the confidence interval or error bounds.

## 2.2 Design Goals

The goals of the IncApprox system are to:

- *Provide application transparency:* We aim to support unmodified applications for stream processing, i.e., the programmers do not have to design and implement application-specific dynamic algorithms or sampling techniques.

- *Guarantee query budget:* We aim to provide an adaptive execution interface, where the users of the system can specify their query budget in terms of tolerable latency/SLAs, desired result accuracy, or the available cluster resources, and our system guarantees the processing within the budget.

- *Improve efficiency:* We aim to achieve high efficiency with a mix of incremental and approximate computing.

- *Guarantee a confidence level:* We aim to provide a confidence level for the approximate output, i.e., the accuracy of the output will remain within an error range. Hence, instead of an exact value of output, IncApprox emits output with a confidence interval as: output $\pm$ error bound.

## 2.3 System Model

Before we explain the design of IncApprox, we present the system model assumed in this work.

### 2.3.1 Programming model

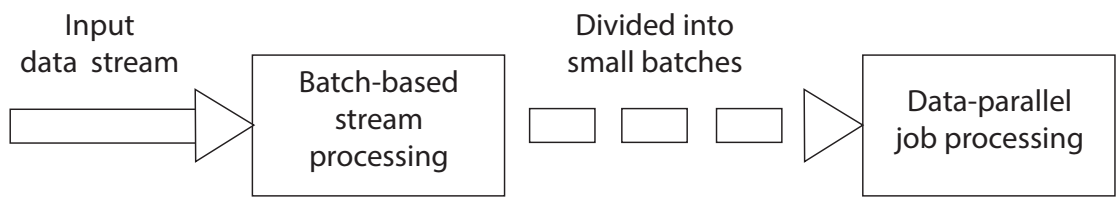

Figure 2.2: Batch-based stream processing

IncApprox supports a *batched streaming processing* programming model. In batched stream processing (see Figure 2.2), the online data stream is divided into small batches or sets of records. Then for each batch, a distributed data-parallel job is launched to produce the output.

As opposed to the trigger-based programming model (see [75] for details), the batched streaming model provides three main advantages: *(i)* it provides simple fault tolerance based on re-computation of tasks, and efficient handling of stragglers using speculative execution; *(ii)* it provides consistent "exact-once" semantics for records processing instead of weaker semantics such as "at least once" or "at most once"; and finally, *(iii)* it provides a unified data-parallel programming model that could be utilized for batch as well as stream processing workflows. Given these advantages, the batched streaming model is widely adopted by many stream processing frameworks including Spark Streaming [6], Flink [2], Slider [23], TimeStream [67], Trident [8], MapReduce Online [32], Comet [48], Kineograph [29], and NOVA [26].

### 2.3.2 Computation model

The computation model of IncApprox for stream processing is *sliding window computations*. In this model (see Figure 2.3), the computation window slides over the input data stream. The new arriving input data items are added to the window and the old data items are dropped from the window as they become less relevant to the analysis.

In sliding window computations, there is a substantial overlap of data items between the two successive computation windows, especially, when the size of the window is large relative to the slide interval. This overlap of unchanged data-items provides an opportunity to update the output incrementally.

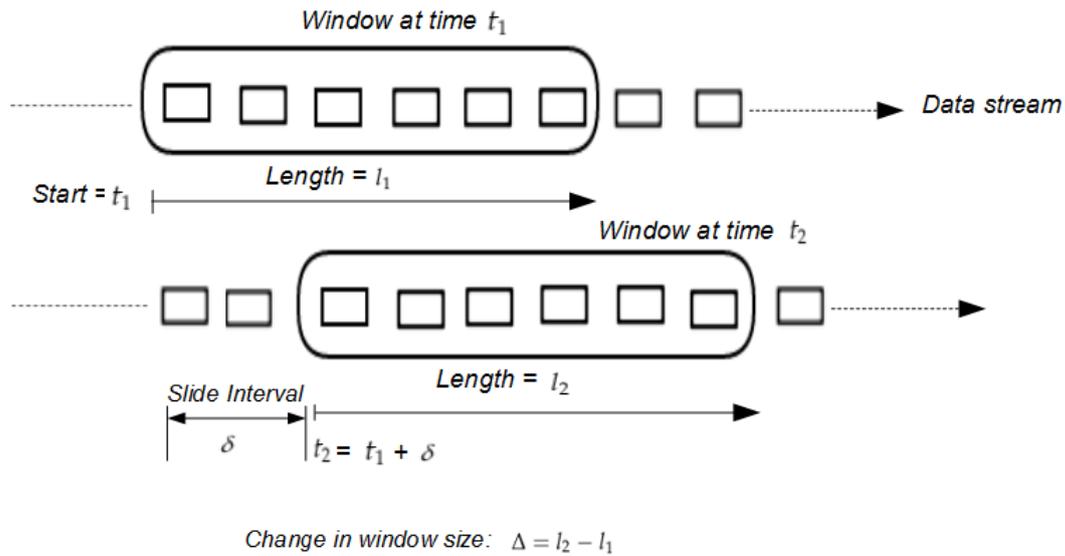

Figure 2.3: Sliding window computation over data stream

### 2.3.3 Assumptions

IncApprox makes the following assumptions. We discuss these assumptions and the different possible methods to enforce them in § 6.

1. We assume that the input stream is stratified based on the source of event, i.e., the data items within each stratum follow the same distribution, and are mutually independent. Here a *stratum* refers to one sub-stream. If multiple sub-streams have the same distribution, they are combined to form a stratum.

2. We assume the existence of a virtual function that takes the user specified budget as the input and outputs the sample size for each window based on the budget.

3. We assume that the memoized results for incremental computation are stored in the way that is fault-tolerant.

Lastly, we assume a time-based window length, and based on the arrival rate, the number of data items within a window may vary accordingly. Note that this assumption is consistent with the sliding window APIs in the aforementioned systems.

## 2.4 Building Blocks

Our system leverages several computational and statistical techniques to achieve the goals discussed in § 2.2. Next, we briefly describe these techniques and the motivation behind our design choices.

### 2.4.1 Stratified sampling

For a realistic rate of execution of the huge amount of streaming data within a sliding window, we perform approximation using samples taken within the window. But the data stream might consist of data from disparate events. So we ensure that every sub-stream is considered fairly to have a representative sample from each sub-stream. For this we use stratified sampling technique [14]. Stratified sampling makes sure that data from every stratum is selected and none of the minorities are excluded. It divides the data within a window into different stratas based on the source of data. Different strata might consist of different number of data items based on how frequently data arrives from the sources. For statistical precision, we use proportional allocation of each sub-stream to the sample [15]. It ensures that the sample size of each sub-stream is in proportion to the size of sub-stream in the whole window. For example, if the data stream consists of two stratas A and B, with 500 items in A and 1000 items in B, and the allowed sample size is 300, then the sample is selected in proportion to data size within each stratum, i.e, here sample would be: 100 items from A, and 200 items from B. So the sample consists of a representative from every strata, and the sample size is proportional to total data size within the window.

### 2.4.2 Self-adjusting computation

We perform incremental sliding window computations by making use of self-adjusting computation [9, 10, 11, 47, 52] to re-use the intermediate results of sub-computations across successive runs of jobs. In this self-adjusting computation technique we maintain a dependence graph between sub-computations of a job, and reuse memoized results for sub-computations that are unaffected by the changed input in the computation window. Since the memoized intermediate results could be re-used this way, we reduce the computation cost and latency.

### 2.4.3 Error estimation

We use error estimation technique [31] for defining a confidence level on the accuracy of the approximated output. This specifies a confidence interval or error bound for the output, i.e., we emit the output in the following form : output $\pm$ error margin. A confidence level along with the margin of error tells how accurate is the approximate output. For example, we express the approximate sum of $n$ values as: $1000 \pm 10$ which means the actual sum lies within the interval 990 to 1010.

# 3 Design

In this section, we present the detailed design of IncApprox.

## 3.1 Algorithm Overview

Algorithm 1 presents an overview of our approach. The user specifies a streaming *query* and the algorithm computes the query as a sliding window computation over the input data stream. The user also specifies a query *budget* for executing the query, which is used to derive the sample size (*sampleSize*) for the window using a cost function (see § 2.3.3 and § 6). The cost function ensures that processing remains within the query budget.

For each window (see Figure 2.3), we first adjust the computation window to the current start time $t$ by removing all old data items from the *window* (*timestamp* $< t$). Similarly, we also drop all old data items from the list of memoized items (*memo*), and the respective memoized results of all sub-computations that are dependent on those old data items.

Next, we read the new incoming data items in *window*. Thereafter, we perform stratified sampling along with proportional allocation on the *window* to select a sample of size provided by the cost function. The cost function ensures that processing remains within the query budget, and proportional stratified sampling ensures that samples from all sub-streams are proportional to the total sub-stream size, and at the same time, no sub-stream is neglected.

Next, we bias the stratified sample to include items from the memoized sample, in order to enable the reuse of memoized results from previous sub-computations. The biased sampling algorithm (detailed in § 3.3) biases samples *specific to each stratum*, to ensure reuse, and at the same time, retain proportional allocation. A proportional stratified sampling in each window is essential since the arrival rate of sub-streams varies for every window. At the same time, in sliding window computations where the window slides by small intervals, there is only a small change in the input based on insertion and deletion from the window (see Figure 2.3), and we can memoize and re-use results of sub-computations whose input remains the same. However, if we re-use *all* memoized results from previous window, the proportional allocation of samples in current window is lost, since the arrival rate of sub-streams may vary largely between windows. Therefore, using biased sampling, we decide *how many memoized items to re-use*, based on the number of items in the current sample from each sub-stream.

Thereafter, on this biased sample, we run the user specified *query* as a data-parallel job *incrementally*. Incremental computation is done by reusing the memoized results for all data items that are unchanged in the window, i.e, in the overlap region of adjacent windows update the output based on the changed (or new) data items. After the job finishes, we memoize all the items in the sample and their respective sub-computation results for reuse for the subsequent windows. The details are covered in § 3.4.

The job provides an estimated output which is bound to a range of error due to approximation. We perform error estimation (as described in § 3.5) to estimate this error bound and define a confidence interval for the result as: *output ± error bound*.

The entire process repeats for the next window, with updated windowing parameters and the sample size. (Note that the query budget can be updated across windows during the course of stream processing to adapt to the user's requirements.)

---

**Algorithm 1 Basic algorithm**
---

    **User input**: streaming *query* and query *budget*
    **Windowing parameters** (see Figure 2.3):
    $t \leftarrow$ start time; $\delta \leftarrow$ slide interval;
    **begin**
        *window* $\leftarrow \emptyset$; // *List of items in the window*
        *memo* $\leftarrow \emptyset$; // *List of items memoized from the window*
        *sample* $\leftarrow \emptyset$; // *Set of items sampled from the window*
        *biasedSample* $\leftarrow \emptyset$; // *Set of items in biased sample*
        **foreach** *window in the incoming data stream* **do**
            // *Remove all old items from window and memo*
            **forall** *elements in the window and memo* **do**
                **if** *element.timestamp* $< t$ **then**
                      *window*.`remove`(*element*);
                      *memo*.`remove`(*element*);
                **end**
            **end**
            // *Add new items to the window*
            *window* $\leftarrow$ *window*.`insert`(*new items*);
            // *Cost function gives the sample size based on the budget*
            *sampleSize* $\leftarrow$ `costFunction`(*budget*);
            // *Do stratified sampling of window (§ 3.2)*
            *sample* $\leftarrow$ `stratifiedSampling`(*window, sampleSize*);
            // *Bias the stratified sample to include memoized items (§ 3.3)*
            *biasedSample* $\leftarrow$ `biasSample`(*sample, memo*);
            // *Run query as an incremental data parallel job for the window (§ 3.4)*
            *output* $\leftarrow$ `runJobIncrementally`(*query, biasedSample*);
            // *Memoize all items & respective sub-computations for sample (§ 3.4)*
            *memo* $\leftarrow$ `memoize`(*biasedSample*);
            // *Estimate error for the output (§ 3.5)*
            *output* $\pm$ *error* $\leftarrow$ `estimateError`(*output*);
            // *Update the start time for the next window*
            $t \leftarrow t + \delta$;
        **end**
    **end**

## 3.2 Stratified Reservoir Sampling

Stratified sampling is a sampling scheme in which the input stream is initially clustered into homogenous disjoint set of strata and from each stratum a random sample is selected. Meanwhile, reservoir sampling selects a uniform random sample of *fixed size* without replacement, from an input stream of unknown size. We perform a combined *stratified reservoir sampling*, adopted from the approach in [14], along with proportional allocation, i.e., we sample the streaming data within a sliding window by stratifying the stream, and applying reservoir sampling within each stratum proportionally. By combin-ing these two techniques, statistical quality of the sample is maintained—as sample from *every stratum* is selected *proportionally*, and a random sample of *fixed size*—given by cost function is selected from the window.

---
**Algorithm 2** Stratified reservoir sampling algorithm
---
   **Require**: *T ← Interval for re-calculation of sub-reservoir size*
   `stratifiedSampling(`*window, sampleSize*`)`
   **begin**
   | *S ← ∅* // *Ordered set of all strata seen so far in window*
   | **forall** *item belonging to stratum $S_i$ in window* **do**
   | | *S*.add($S_i$); // *Add new stratum seen to S*
   | | *i ← Index of stratum $S_i$* ;
   | | // *Fill reservoir until sampleSize is reached*
   | | **if** ($\sum_{h=1}^{|S|} |sample[h]|$) < *sampleSize* **then**
   | | | *sample*[*i*].add(*item*); // *Add item to its sub-reservoir*
   | | **end**
   | | **else**
   | | | **if** *T interval is passed* **then**
   | | | | **forall** $S_i$ *in S* **do**
   | | | | | *i ← Index of stratum $S_i$* ;
   | | | | | // *Compute new sub-reservoir size using Equation 3.1*
   | | | | | *newSize*[*i*] ← *sample*[*i*].computeSize();
   | | | | | **if** *newSize*[*i*] ≠ |*sample*[*i*]| **then**
   | | | | | | *c ← newSize*[*i*] − |*sample*[*i*]|;
   | | | | | | // *Do Adaptive Reservoir Sampling*
   | | | | | | *sample*[*i*] ← `ARS`(*c, sample*[*i*], $S_i$);
   | | | | | **end**
   | | | | | **else**
   | | | | | | // *Do Conventional Reservoir Sampling*
   | | | | | | *sample*[*i*] ← `CRS`(*item, sample*[*i*], $S_i$);
   | | | | | **end**
   | | | | | // *Skip items in window, if seen by ARS or CRS*
   | | | | | `skipItemsSeen();` // *Details omitted*
   | | | | **end**
   | | | **end**
   | | | **else**
   | | | | // *Until T, do Conventional Reservoir Sampling*
   | | | | *sample*[*i*] ← `CRS`(*item, sample*[*i*], $S_i$);
   | | | **end**
   | | **end**
   | **end**
   **end**

---

The stratified reservoir sampling algorithm (described in Algorithm 2) uses a fixed size reservoir. The size of the reservoir is equal to the sample size given by cost function. The algorithm allocates the space in the reservoir proportionally to the samples from each stratum, based on number of items seen so far in the corresponding stratum. While we move forward through the window for sampling, the arrival rate of items in each stratum may change due to the stream of data coming in. Hence the proportional allocation must be updated. Therefore, the algorithm re-allocates the space in the reservoir periodically to ensure proportional allocation. Thereafter, based on this re-allocation, we adapt the algorithm to use an adaptive reservoir sampling (ARS) [15] for those strata whose sub-reservoir sizes are changed, and conventional reservoir sampling (CRS) [14] for those strata whose sub-reservoir sizes are unchanged. (Let reservoir consists of a group of sub-reservoirs, each for storing sample from each stratum). ARS ensures that we periodically adjust the proportional allocation (based on the arrival rate), and CRS ensures randomness in sampling technique. Once the sub-reservoir's proportional allocation is handled using ARS, the sampling technique switches back to CRS, until the next re-allocation interval.

Algorithm 2 works as follows: For each *item* seen in a window, if the stratum of the item is newly seen, then we add it to the set of strata seen so far. Initially, we fill the reservoir of sample until it is full. Here the reservoir is a store for our stratified sample *'sample'*, and can be considered as a group of sub-reservoirs of different strata such that: $|sample| = \sum_{i=0}^{|S|-1} |sample[i]|$ where $S$ is the ordered set of all strata seen so far in the window, and *sample[i]* is the sub-reservoir of the sample from the $i^{th}$ stratum. We fill the reservoir by adding each *item* to its corresponding sub-reservoir, based on the stratum to which the *item* belongs.

Once the reservoir is full, then until a pre-decided periodical time interval $T$ to re-allocate sub-reservoir sizes, we proceed with a conventional reservoir sampling (CRS). In CRS technique, for each of the further items seen in each stratum $S_i$, we decide with a probability $\frac{|sample[i]|}{|S_i|}$ whether to accept or reject the item, i.e., all items in a stratum have equal probability of inclusion [14]. If the item is accepted, then we replace a randomly selected item in the corresponding sub-reservoir with the accepted item. After $T$ interval of time, we re-allocate the sub-reservoir sizes of each stratum, to ensure proportional allocation. This $T$ interval determines how frequently proportional allocation is verified. Thus, $T$ is selected based on frequency of change in the arrival rate in each stratum (since change

Chapter 3.  Design 3.2. Stratified Reservoir Sampling

**Algorithm 3 Subroutines for the stratified sampling algorithm**

*Let incomingItems[ ] represent incoming items seen when moving forward through window*
```
ARS(c, sample[i], S_i)
begin
    if c > 0 then
        // Add c items to sample[i] from incoming items belonging to S_i
        ∀j ∈ {0, ..., c − 1} : sample[i].add(incomingItems[S_i].get(j));
    end
    else
        // Evict random c items from sample[i]
        ∀j ∈ {0, ..., c − 1} :
        // random(a, b) gives a random number between [a, b]
        sample[i].remove(random(0, |sample[i]| −1));
    end
end
CRS(item, sample[i], S_i)
begin
    p ← |sample[i]| / |S_i|; // Probability of replacement
    // Replace a random item from sample[i] with item, using probability p
    sample[i].replace(sample[i][random(0, |sample[i]| −1)],
    item, p);
end
```

in arrival rate changes proportional allocation), by counting the number of items of each stratum per time unit at the stream aggregator. First, after interval $T$, we compute the size of sub-reservoir to be allocated to each $i^{th}$ stratum at current time $t'$. It is computed proportional to the total number of items seen so far in the corresponding stratum within the window, using the equation:

$$|sample[i](t')| = sampleSize * \frac{|S_i|}{k} \qquad (3.1)$$

where *sampleSize* is the total size allocated to reservoir, $|S_i|$ is the number of items seen so far in the stratum $S_i$ and $k$ is the total number of items seen so far in the window. For e.g., suppose samples selected from each strata $S_1, S_2 ..., S_n$ are stored in sub-reservoirs $|sample[1], |sample[2]|, ... |sample[n]|$ then, $|sample[i]|$ is proportional to $|S_i|$ where $S_i$ is the $i^{th}$ strata and $sample[i]$ is its corresponding sub-reservoir.  Thereafter, if the re-allocated sub-reservoir size $|sample[i](t')|$ at current point of time $t'$ is different from the previously adjusted sub-reservoir size (i.e., if there is any change in sub-reservoir size), we proceed with ARS—to adapt according to this change in size (described in Al-gorithm 3) as follows: When sub-reservoir size of $S_i$ has increased by $c$, then from the incoming stream, we insert $c$ items that belong to stratum $S_i$, to the corresponding sub-reservoir $sample[i]$. If the sub-reservoir size has decreased by $c$, we evict $c$ number of items from the sub-reservoir. This ensures that proportional allocation is retained.

Here, we assume that since $|sample[i]|$ has been filled from $|S_i|$ items, then the room for $c$ items could be filled from incoming $c * \frac{|S_i|}{|sample[i]|}$ items. We also make sure that, while proceeding with ARS for any $i^{th}$ stratum $S_i$, if we see any new stratum, then it is updated in the set of strata $S$ so that no stratum is left out, and the sub-reservoir allocation of every stratum is adjusted accordingly in the next re-allocation.

If the re-allocated sub-reservoir size of a stratum is unchanged, we proceed with CRS for the stratum i.e, for each item seen, we decide with a probability $\frac{|sample[i]|}{|S_i|}$ whether to accept or reject item. If the item is accepted, then we replace a randomly selected item in the sub-reservoir with the accepted item.

We perform stratified reservoir sampling until the window terminates and the resulting stratified sample consists of samples from each stratum, proportional to the size of corresponding stratum seen in the whole window.

## 3.3 Biased Sampling

### 3.3.1 Biased sampling

After stratified reservoir sampling, we perform incremental computation to re-use the memoized results from previous computations. However, if we reuse *all* memoized results from the previous window, the proportional allocation is lost, since proportions in different windows may vary due to difference in the arrival rate of sub-streams. Here face three issues such as *(i) number of items* sampled from each stratum in current window might vary from number of items memoized from that stratum from previous window since the arrival rate of each sub-stream fluctuates; *(ii)* even if number of items memoized and number of items present in the sample of each stratum are the same, the *items* memoized might differ from items sampled, due to the randomness in reservoir sampling, and hence we cannot re-use all the memoized results; and *(iii)* sample from current window might already include some of the memoized items, but not all of them.

In order to solve these problems, we use biased sampling technique. Biased sampling enables result reuse by including memoized data items in the sample, but at the same time, ensures that the proportional allocation of samples from each stratum is retained. We solve each of the above mentioned issues by performing biased sampling *separately for each stratum* as follows: We solve the above mentioned issues as follows - for solving *(i)*: if the number of items in the sample from current window is greater than the number of memoized items, then we re-use all memoized items, and if it is lesser, we re-use only the number of items required in the current sample and neglect the extra memoized items - this guarantees the proportional allocation even after re-use. We update the set of samples by replacing its items with the memoized items accordingly, ensuring proportional allocation; for solving *(ii)*: since we give priority to the memoized items, we replace the items in the current sample with the memoized items; and for solving *(iii)*: we make sure that while replacing items in current sample using memoized items(as described in previous steps), there is no duplicate items in the current sample. For this, we use a hash based data structure e.g, Hashset. After this first phase of incremental computation, we get a final sample which includes all essential memoized items and new items sampled from current window based on its sub-sample sizes.

Algorithm 4 describes our biased sampling algorithm. In this algorithm, we bias the sample from each stratum separately. Note that here, "memoized items" and "sample size" are specific to each stratum. The algorithm works as follows: If the number of memoized items $x$ is greater than or equal to the sample size $y$, then we create a biased sample with only $y$ items from the memoized list, and neglect the extra memoized items. If the number of memoized items is less than the sample size, then we give priority to memoized items and create a biased sample with all memoized items first, and later we add more items to this biased sample from the stratified sample until the size of biased sample becomes equal to the size of stratified sample. This ensures proportional allocation. However, some of the memoized items in *memo* might be already in the stratified sam-

**Algorithm 4 Biased sampling algorithm**

```
biasSample(sample, memo)
begin
    S ← sample.getAllStrata(); // Set of all strata in sample
    foreach i^th stratum S_i in S do
        x ← memo[i].size(); // no. of items memoized from S_i
        y ← sample[i].size(); // no. of items in sample from S_i
        biasedSample[i] ← ∅; // List of items in biased sample from S_i
        if x ⩾ y then
            // Add y items from memo[i] to biasedSample[i] to enable re-use
            ∀j ∈ {0, ..., y − 1} : biasedSample[i].add(memo[i].get(j));
        end
        else
            // First add x items from memo[i] to biasedSample[i]
            ∀j ∈ {0, ..., x − 1} : biasedSample[i].add(memo[i].get(j));
            // Fill the remaining (y − x) items from the stratified sample
            int j = 0;
            while (biasedSample[i].size() < y) do
                biasedSample[i].add(sample[i].get(j));
                j++;
            end
        end
    end
end
```

ple, and this might cause duplicates in the biased sample. Therefore, in practice, we use a data structure such as a HashSet for storing *biasedSample* to remove duplicates automatically. Finally, we get a biased sample which includes all essential memoized items as well as stratified samples based on the arrival rate, thus ensuring both reuse and proportional allocation.

### 3.3.2 Precision and accuracy in biased sampling

An estimated result is *precise* if similar results are obtained with repeated sampling, and it is *accurate* if the estimated result is closer to the true result (a precise result doesn't necessarily be accurate always) [56]. Our stratified sample is precise than a random sample since it considers every stratum, and uses proportional allocation. Accuracy of a stratified sample is more if *(i)* different strata have major differences and *(ii)* within each stratum, there is homogeneity [56]. Based on our assumptions in 2.3.3, our stratified sample is ac-curate since different stratum have different distribution, and items within each stratum follow the same distribution (homogenous).

We bias the sample from each stratum separately, thus preserving the statistics of stratified sampling, i.e., after the bias, the biased sample still consists of items from each stratum in the same proportional allocation obtained from stratified sampling. Further, even though the items selected within a stratum are biased to include memoized items which belong to the same stratum, since the items follow same distribution, there is little difference between items within a stratum. Thus our bias sampling technique is as precise and accurate as how the stratified sample is, provided the assumptions hold.

## 3.4 Run Job Incrementally

Next, we run the user-specified streaming query as an *incremental* data-parallel job on the biased sample (derived in § 3.3). For that, we make use of self-adjusting computation [9, 10, 47].

In self-adjusting computation, the computation is divided into sub-computations, and a dynamic dependence graph is constructed to record dependencies between these sub-computations. Formally, a Dynamic Dependence Graph $DDG = (V, E)$ consists of nodes ($V$) representing sub-computations and edges ($E$) representing data and control dependencies between the sub-computations. Thereafter, a change propagation algorithm is used to update the output by propagating the input changes through the dependence graph. The change propagation algorithm identifies a set of sub-computations that directly depend on the changed data and re-executes those sub-computations. This in turn leads to re-computation of other data-dependent sub-computations. Change propagation terminates when all transitively dependent sub-computations are re-computed. For all the unaffected sub-computations, the algorithm reuses memoized results from previous runs without re-computation. Lastly, results for all re-computed (or newly computed) sub-computations are memoized for the next incremental run.

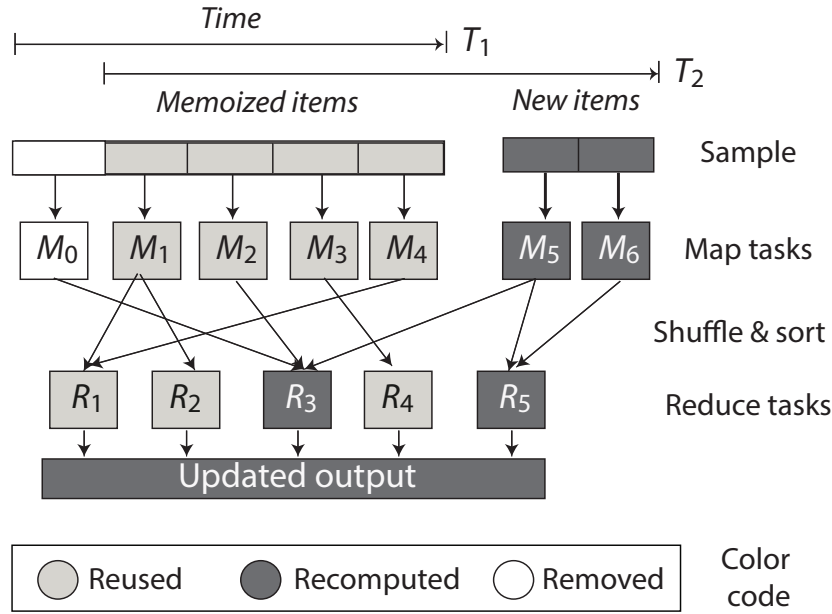

Figure 3.1: Run data-parallel job incrementally

Next, we illustrate the application of self-adjusting computation to a data-parallel job based on the MapReduce model [34]. (Note that our implementation is based on Spark Streaming [6], which is a generic extended version of MapReduce.) Figure 3.1 shows the dependence graph built based on the data-flow graph of the MapReduce model. The data-flow graph is represented by a DAG, where *map* and *reduce* tasks represent nodes (or sub-computations) in the dependence graph, and the directed edges represent the dependencies between these tasks. For an incremental run, we launch *map* tasks for the newly added data items in the sample ($M_5$ and $M_6$), and reuse the memoized results for the *map* tasks from previous runs ($M_1$ to $M_4$). The output of the newly computed *map* tasks invalidates the dependent *reduce* tasks ($R_3$ and $R_5$). However, all *reduce* tasks that are unaffected by the changed input can simply reuse their memoized results without re-computation ($R_1$, $R_2$, and $R_4$). Lastly, we memoize the results for all *freshly* executed tasks for the next incremental run. Note that the items removed from the window also act as the input change (e.g., $M_0$), and sub-computations dependent on the removed items are also re-computed (e.g., $R_3$).

## 3.5 Estimation of Error Bounds

In order to provide a confidence interval for the approximate output, we estimate the error bounds due to approximation.

### 3.5.1 Approximation for aggregate functions

Aggregate functions require results based on all the data items or groups of data items in the population. But since we compute only over a small sample from the population, we get an *estimated* result based on the weightage of the sample.

Consider an input stream $S$, within a window, consisting of $n$ disjoint strata $S_1$, $S_2$ ..., $S_n$, i.e., $S = \sum_{i=1}^{n} S_i$. Suppose the $i^{th}$ stratum $S_i$ has $B_i$ items and each item $j$ has an associated value $v_{ij}$. Consider an example to take sum of these values, across the whole window, represented as $\sum_{i=1}^{n}(\sum_{j=1}^{B_i} v_{ij})$. To find an approximate sum, we first select a sample from the window based on stratified and biased sampling as described in § 3, i.e., from each $i^{th}$ stratum $S_i$ in the window, we sample $b_i$ items. Then we estimate the sum from this sample as: $\hat{\tau} = \sum_{i=1}^{n}(\frac{B_i}{b_i} \sum_{j=1}^{b_i} v_{ij}) \pm \epsilon$ where the error bound $\epsilon$ is defined as:

$$\epsilon = t_{f, 1-\frac{\alpha}{2}} \sqrt{\widehat{Var}(\hat{\tau})} \tag{3.2}$$

Here, $t_{f, 1-\frac{\alpha}{2}}$ is the value of the t-distribution (i.e., *t-score*) with $f$ degrees of freedom and $\alpha = 1 -$ confidence level. The degree of freedom $f$ is expressed as:

$$f = \sum_{i=1}^{n} b_i - n \tag{3.3}$$

The estimated variance for sum, $\widehat{Var}(\hat{\tau})$ is represented as:

$$\widehat{Var}(\hat{\tau}) = \sum_{i=1}^{n} B_i * (B_i - b_i) \frac{s_i^2}{b_i} \tag{3.4}$$

where $s_i^2$ is the population variance in the $i^{th}$ stratum. Since the bias sampling is such that the statistics of stratified sampling is preserved, we use the statistical theories [70] for stratified sampling to compute the error bound.

Currently, we support error estimation only for aggregate queries. For supporting queries that compute extreme values, such as minimum and maximum, we can make use of extreme value theory [31, 54] to compute the error bounds.

### 3.5.2 Error bound estimation

Accuracy of an estimated result is indicated by a confidence interval. For example, a 95% confidence interval is explained as: if we take samples from our population over and over again, and construct a confidence interval using our procedure for each possible sample, we expect 95% of the resulting intervals to include the true value of the population parameter[57]. For error bound estimation, we first identify the sample statistic used to estimate a population parameter, e.g., *sum*, and we select a desired confidence level, e.g., 95%. In order to compute the margin of error $\epsilon$ using t-score as given in Equation 3.2, the sampling distribution must be nearly normal. The Central Limit Theorem (CLT) states that when the size of sample is sufficiently large ($>= 30$), then the sampling distribution of a statistic approximates to *normal distribution*, regardless of the underlying distribution of values in the data [70]. Hence, we compute t-score using a t-distribution calculator [60], with the given degree of freedom $f$ (see Equation 3.3), and cumulative probability as $1 - \alpha/2$ where $\alpha = 1 -$ confidence level [56]. Thereafter, we estimate the variance using the corresponding equation for the sample statistic considered (for *sum*, the Equation is 3.4). Finally, we use this t-score and estimated variance of the sample statistic and compute the margin of error using Equation 3.2.

# 4 Implementation

We implemented INCAPPROX based on the Apache Spark Streaming framework [6]. Figure 4.1 presents the high-level architecture of our prototype. Our system uses Apache Kafka [5] as stream aggregator that collect streams from multiple sources and generate an aggregated data stream. INCAPPROX processes this streaming data from kafka and provides an approximate result.

In this section, we first give a brief necessary background on Apache Kafka and Apache Spark streaming, and next, we present the design details of the different modules in INCAPPROX.

## 4.1 Background

### 4.1.1 Apache Kafka

Kafka [5] is a messaging-based log aggregator with the combined benefits of traditional log aggregators and messaging systems. It is based on publish/subscribe model and is a distributed, high-throughput messaging system. It provides a simple API similar to messaging systems and allow applications to consume events in real-time. In kafka, a stream of messages that belongs to a particular type is defined by *topic*. Producers can produce messages and publish to a particular topic. These published messages are stored to a set servers called brokers. Consumers can subscribe to one or more topics from the brokers. Consumers consume the messages using a pull model i.e, by pulling data from the brokers.

The messages produced by a kafka producer is published to a topic. For subscribing to a topic, a consumer first creates one or more message streams for the topic. Then the messages published to that topic will be evenly distributed into these sub-streams by kafka. Kafka also provides an iterative interface over the continual stream of messages being produced for each message stream. The consumer iterates over every message in the stream and processes the payload of the message.

### 4.1.2 Apache Spark Streaming

Spark Streaming [6] is a scalable and fault-tolerant distributed stream processing framework. It offers batched stream processing APIs (as described in § 2.3), where a streaming computation is treated as a series of batch computations on small time intervals. For each interval, the received input data stream is first stored on a cluster's memory and a distributed file system such as HDFS [4] or Tachyon [53]. Thereafter, the input data is processed using Apache Spark [74], a distributed data-parallel job processing framework similar to MapReduce [34] or Dryad [50].

Spark Streaming is built on top of Apache Spark, which provides a high-level abstraction called Resilient Distributed Datasets (RDDs) [74] for distributed data-parallel computing. An RDD is an immutable and fault-tolerant collection of elements (objects) that is distributed or partitioned across a set of nodes in a cluster. Spark Streaming extends the RDD abstraction by introducing the DStreams APIs [75], which is a sequence of RDDs arrived during a time window.

## 4.2 IncApprox implementation

Our implementation builds on the Spark Streaming APIs to implement the approximate and incremental computing mechanisms. At a high-level (see Figure 4.1), the input data stream is split into batches based on a pre-defined interval (e.g., one second). Each batch is defined as a sequence of RDDs. Next, the RDDs in each batch are sampled by the sampling module, with a sampling rate computed from the query budget using the virtual cost function. The sampled RDDs are inputs for the incremental computation module. In this module, the sampled RDDs are processed incrementally to provide the query result to the user. Finally, the error is estimated by the error estimation module. We next explain the details of the different modules of INCAPPROX.

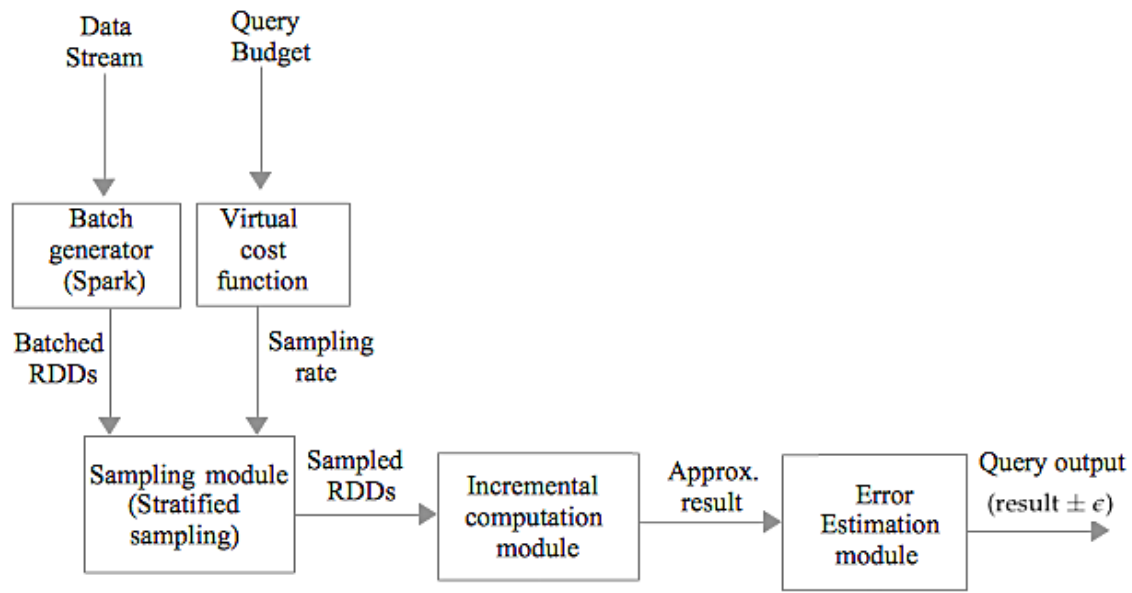

Figure 4.1: Architecture of INCAPPROX prototype

### 4.2.1 Sampling module

Our sampling module implements the approximation mechanism as described in § 3.

Input data for sampling comes from kafka which act as the stream aggregator. Multiple kafka producers publish data to specific topics. This data is stored in the kafka brokers. Our system uses a single consumer that consumes data from these different topics. For this, the consumer first creates a message stream( i.e, a ReceiverInputDStream) to receive the data for the topics it wants to subscribe. Into this receiver stream, kafka distributes the messages published to the subscribed topics. This data stream acts as the input to our sampling module.

Our sampling module performs stratified sampling along with proportional allocation. For implementing stratified sampling algorithm, we adapt the sampling methods available on Spark. For example, for extracting a stratified sample, we use *sampleByKey()* on RDD's of key-value pairs where key represents the sub-stream ( eg; a label of sub-stream) and value represents an item that belongs to the sub-stream. This allows us to sample the data in each of the sub-streams separately. Spark's *sampleByKey()* also allows to specify the sampling fraction required from each sub-stream to maintain proportional allocation. Our sampling module thus samples data from each stratum without neglecting the minority groups.

### 4.2.2 Incremental computation module

Our incremental computation module implements the self-adjusting computation mechanism as described in § 3.4. To implement this component, we reuse the caching mechanism available in Spark to memoize the intermediate results for the tasks. For the reduction operations, we adapt a windowing operation in Spark Streaming, namely *reduceByKeyAndWindow()*. This function allows us to use a time-based window frame to reduce over. For this, we provide a reduce function and an inverse reduce function. Then for each iteration within the window frame, Spark will reduce the new data and "un-reduce" the old. Finally, the dependence graph is maintained at Spark's job controller.

### 4.2.3 Error estimation module

Finally, the error estimation module calculates the error bounds for the output. The error estimation algorithm described in § 3.5 can be easily implemented by re-using the error calculation functions provided by *Apache Common Math* library [60]. Further details on this can be found in our paper [51].

In general, our modifications in Spark Streaming are fairly straightforward, and could easily be adapted to other batched streaming processing frameworks (described in § 2.3). More importantly, we support unmodified applications since we did not modify the application programming interface.

# 5 Evaluation

In this section, we present a micro-benchmarks based evaluation (§ 5.1) of INCAPPROX.

## 5.1 Micro-benchmarks

For analyzing the effectiveness of memoization in improving the result reuse rate, we evaluate INCAPPROX using a simulated data stream. In particular, our evaluation analyzes the impact of varying four different parameters, namely, sample size, slide interval, window size, and arrival rate for sub-streams.

We generated a synthetic data stream with three different sub-streams. Each sub-stream is generated with an independent Poisson distribution and different mean arrival rates. For the first three experiments, i.e., to analyze the impact of sample size, slide interval, and window size on memoization, we generated three different sub-streams with a mean arrival rate of 3 : 4 : 5 data items per unit time respectively. To analyze the impact of the fluctuating arrival rate of events, we generated two sub-streams with fluctuating arrival rates, and kept the third sub-stream with a constant arrival rate, for a comparative analysis.

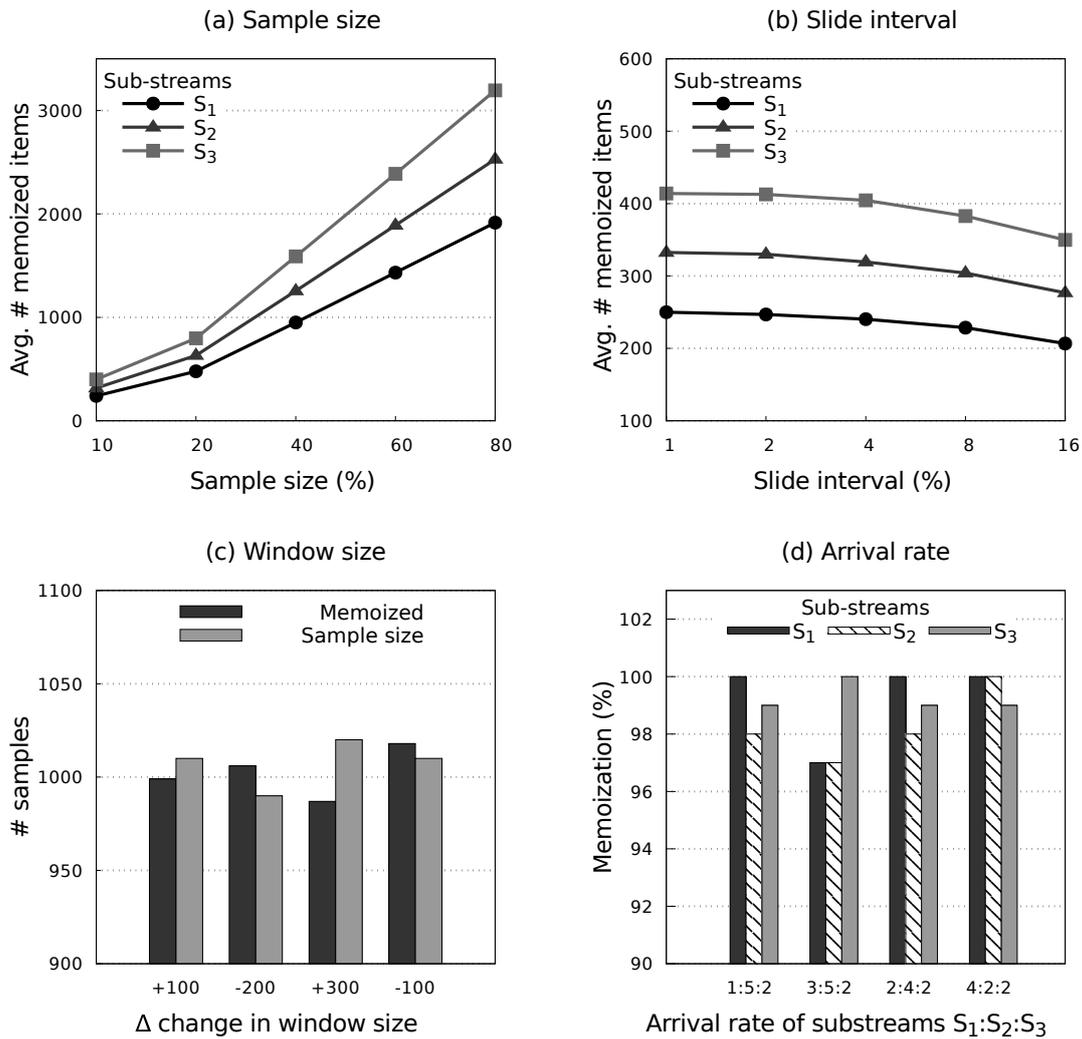

Figure 5.1: Effect of various parameters such as sample size, slide interval, window size and arrival rate on memoization

### 5.1.1 What is the effect of varying sample sizes?

We first study the effect of varying sample sizes on memoization by applying our algorithm (described in § 3) to the synthetic data stream. For the experiment, we keep the other parameters—window size and slide interval—fixed. We measure the average number of memoized items from each sub-stream $S_1$, $S_2$, $S_3$ with different arrival rates $3:4:5$ respectively, by varying the total sample size.

Figure 5.1 (a) shows our measurements with a fixed window size of $10,000$ items, 4% slide interval (i.e., 400) and varying sample sizes (on x-axis): 10%, 20%, 40%, 60% and 80% of the window size. We observe that the average number of data items memoized is directly proportional to the sample size and the arrival rate. When the sample size increases, the average number of data items memoized increases constantly because more items from the previous window is available for memoization. We also observe a higher memoization rate for sub-streams with higher arrival rates, the reason being a proportional allocation of sub-sample sizes.

### 5.1.2 What is the effect of varying slide intervals?

Next, we evaluate the impact of varying slide intervals on memoization with constant window and sample sizes. We measure the average number of items memoized from each sub-stream with different slide intervals.

Figure 5.1 (b) shows our measurements with a fixed window size of $10,000$ and sample size of 10% window size (i.e., 1000), but varying slide intervals (on x-axis): 1%, 2%, 4%, 8%, and 16% of the window size. We observe that when the slide interval is 1%, our algorithm memoizes an average of 99.5% of total samples, which greatly improves the reuse rate, and thus, leads to higher efficiency. As evident from the plot, when the slide interval increases, the percentage of memoized items decreases, because larger slides allow fewer samples to reuse from the previous window. We also repeated the experiments with different window sizes, but observed very similar results. Thus, the results illustrate that smaller slides (which is the usual case for an incremental workflow) allow higher memoization and thus higher result reuse.

### 5.1.3 What is the effect of varying window sizes?

Next, we evaluate the impact of varying window sizes on memoization. This is done by a comparison of sample size and memoization rate for each window size i.e., we measure the number of items in a sample and the number of items memoized from the previous window. Then we analyze the reuse rate based on this measurement. We begin our experiment with a window of $10,000$ items, and then increase/decrease the window size, e.g., we first increase the window size by 200, then decrease it by 100, etc., to get windows of varying sizes.

Figure 5.1 (c) shows our measurements, with a fixed 2% slide interval and 10% sample size for each corresponding window size. The x-axis represents $\Delta$, i.e., the change in window size between two adjacent windows (see Figure 2.3). The figure illustrates that whenever the window size decreases (*i.e., $\Delta$ is negative*), memoized samples are more than the samples needed in the current window. For example, when $\Delta$ is $-100$, sample size is 1010 and we have 1017 memoized items from the previous window i.e., decreasing window size can allow a 100% re-use rate, provided the slide interval is considerably low (here 2%). The figure also depicts that when window size increases (*i.e, $\Delta$ is positive*), the sample size is higher than the number of memoized items from the previous window, and the larger the increase in the window size, the larger is the difference between samples needed and memoized. This implies a lesser result reuse rate.

### 5.1.4 What is the effect of fluctuating arrival rates?

Lastly, we evaluate the effect of fluctuating arrival rate of sub-streams. As mentioned earlier, we generated two sub-streams, each with fluctuating arrival rates, and a third sub-stream with a constant arrival rate for the analysis. We measure the percentage of items memoized from each sub-stream.

Figure 5.1 (d) depicts the memoization based on fluctuating arrival rates, for a fixed window of 10,000 items and sample size of 10%. The x-axis shows the arrival rate for the three substreams $S_1$, $S_2$, and $S_3$. The figure illustrates that memoization is inversely proportional to the arrival rate. For example, for sub-stream $S_1$, when the arrival rate increases from 1 to 3, the percentage of memoization decreases, because the sample size gets higher due to proportional allocation, but memoized items available are lesser. When $S_1$'s arrival rate decreases from 3 to 2, we observe that the memoization increases since we have more items memoized from the previous window. Sub-stream $S_2$ also depicts similar behaviour. However we notice that even though arrival rate is constant for the third sub-stream, its memoization rate differs relative to the other two sub-streams since we use a proportional allocation of sample sizes. The figure illustrates that in spite of the fluctuations in arrival rates, INCAPPROX has a memoization rate greater than 97%.

# 6 Discussion

The design of INCAPPROX is based on three assumptions (see § 2.3.3). Solving these assumptions is beyond the scope of this thesis; however, in this section, we discuss some of the approaches that could be used to meet our assumptions.

## 6.1 Stratification of sub-streams

Currently we assume that sub-streams are stratified, i.e., the data items of individual sub-streams have the same distribution. However, it may not be the case. Next, we discuss two alternative approaches, namely bootstrap [38, 39, 65] and a semi-supervised learning algorithm [59] to classify evolving data streams.

Bootstrap [38, 39, 65] is a non parametric re-sampling technique used to estimate parameters of a population. It works by randomly selecting a large number of bootstrap samples with replacement and with the same size as in the original sample. Unknown parameters of a population can be estimated by averaging these bootstrap samples. We could create such a bootstrap-based classifier from the initial reservoir of data, and the classifier could be used to classify sub-streams. Alternatively, we could employ a semi-supervised algorithm [59] to stratify a data stream. This algorithm manipulates both unlabeled and labeled data items to train a classification model, which can be used to stratify the data stream. Such an approach is preferred over supervised algorithms as most items in a data stream may not be labelled.

## 6.2 Virtual cost function

Secondly, we assume that there exists a virtual function that computes the sample-size based on the user-specified query budget. The query budget could be specified as either available computing resources or latency requirements. We suggest two existing approaches—Pulsar [16] and resource prediction model [41, 58]—to design such a virtual function for given computing resources and latency requirements, respectively.

Pulsar [16] is a system that allocates resources based on tenants' demand, using a multi-resource token bucket. It provides a workload independent guarantee using a pre-advertised cost model, i.e., for each appliance and network, it advertises a virtual cost function that maps a request to its cost in tokens. We could adopt a similar cost model as follows: An "item", i.e., *a data block* to be processed, could be considered as a request and "amount of resources" needed to process it could be the cost in tokens. Since the resource budget gives total resources (*here tokens*) to be used, we could find the number of items, i.e., the sample size, that can be processed using these resources, ruling out faults and stragglers.

To find the sample-size for a given latency budget, we could use a resource prediction model based on performance metrics and QoS parameters in SLAs. Such a model could analyze the diurnal patterns of resource usage [27], e.g., off-line predictions based on pre-recorded resource usage log or predictions based on statistical machine learning [41, 58], to predict the future resource requirements based on workload and latency. Once we get the resource requirement for a latency budget using this model, we could find the sample-size needed by using the above suggested method similar to Pulsar.

## 6.3 Fault tolerance

Our current algorithm does not take into account the failure of nodes in the cluster where memoized results are stored. We discuss three different approaches that could be adopted for fault tolerance if memoized results are unavailable: *(i)* we could continue processing the window without using any memoized items, albeit with lower efficiency; *(ii)* we could use a similar approach for fault tolerance as provided in Spark [74], where the lineage of memoized RDDs is used to recompute only the lost RDD partitions; That is, if any partition of an RDD is lost due to a worker node failure, then that partition could be re-computed from the original fault-tolerant dataset using the lineage; *(iii)* we could make use of underlying distributed fault tolerant file-systems (HDFS [4]) to *asynchronously* replicate the memoized results. or the RDD's of only those memoized results which are frequently queried. This is based on the heuristic that if a result has been frequently queried, then it is likely to be queried in the future too.

# 7 Related Work

INCAPPROX builds on two computing paradigms, namely, incremental and approximate computing. In this section, we survey the techniques proposed in these two paradigms of computing.

## 7.1 Incremental computation

Since modifying the output of computation incrementally is asymptotically more efficient than re-computing everything from scratch, incremental computation is an active area of research for "big data" analytics.

Incremental computation on large-scale data sets can be broadly divided into two categories: non-transparent and transparent approaches. Non-transparent approach was introduced by earlier big data systems for incremental computation, where the programmer is asked to implement an efficient incremental-update mechanism. Examples of non-transparent systems include Google's Percolator [64], and Yahoo!'s CBP [55]. Using percolator [64], the programmer requires to write programs in an event-driven programming model. Percolator structures an application as a series of observers which are triggered by the system when-ever user-specified data changes. Yahoo!'s CBP (continuous bulk processing) [55] provides "bulk incremental processing" on traditional MapReduce platforms by running MapReduce jobs on new data every few minutes. It introduces new primitives to store and reuse prior state for incremental processing. In particular, loop-back flows redirect the output of a stage as the input for subsequent runs. Even though these systems have brought performance benefits, they require the programmer to devise a dynamic algorithm in order to efficiently process data in an incremental manner. These algorithms can be difficult to design, implement and maintain even for simple problems. Moreover, most dynamic algorithms are designed for sequential computing model, and cannot be easily adopted to parallel and distributed setting, which is commonly used for analyzing large datasets.

To overcome the limitations of the aforementioned systems, researchers proposed transparent approaches for incremental computation [71, 72, 73]. Examples of transparent systems include Incoop [18, 19], Comet [48], NOVA [26], DryadInc [66] and Slider [20, 23]. These systems leverage the underlying data-parallel programming model such as MapReduce [34] or Dryad [50] for supporting incremental computation.

Incoop [18, 19] is a generic MapReduce framework for transparent incremental computations on streaming data without requiring the programmers to design dynamic algorithms for incremental computing. It transparently executes the Map-Reduce programs in an incremental manner i.e, whenever the input data changes, computations automatically updates the output incrementally by re-using the intermediate results of previous computations. For this, Incoop relies on memoization by performing a stable partitioning of the input and by reducing the granularity of tasks to maximize result-reuse. It also identifies the short-comings of task-level memoization approaches, and address them using several novel techniques such as a storage system to store the input of consecutive runs, a contraction phase that make the incremental computation of the reduce tasks more efficient, and a scheduling algorithm for Hadoop that is aware of the location of previously computed results.

Comet [48] is another system that uses transparent incremental computations. It introduces batched stream processing (BSP) and models data as a stream with queries triggered upon bulk appends to the stream. Comet exploits temporal and spatial correlations of recurring queries by defining the notion of query series. Executions within a query series will automatically leverage the intermediate results of previous executions of the same query. It also aligns multiple queries to execute together when new bulk updates occur.

Another system called NOVA [26] proposed by Yahoo! is designed for the incremental execution of Pig programs upon continually-arriving data. It introduces a new layer called work flow manager, on top of Hadoop framework, which identifies computations affected by incremental changes and produce necessary update functions to run on top of Hadoop framework.

All of the above systems propose distinct approaches of incremental computing. Our work builds on transparent big data systems for incremental computation. In particular, we leverage the advancements in self-adjusting computation [9, 17] to improve the efficiency of incremental computation. In contrast to the existing approaches, our approach extends incremental computation with the idea of approximation, thus further improving the performance and throughput for applications.

## 7.2 Approximate computation

Approximate Query Processing (AQP) for decision support in relational databases has been the subject of extensive research. Approximation techniques such as sampling [14, 28], sketches [33], and online aggregation [49] have been well-studied over the decades in the context of traditional (centralized) database systems. Some of th widely used sampling based approximate query processing systems include STRAT [28], SciBORQ [69], ApproxHadoop [43]and BlinkDB [12, 13].

STRAT [28] uses stratified sampling technique for approximation. It introduces the concept of fundamental regions (FRs). For a given relation R and workload W, consider partitioning the records in R into a minimum number of regions $F = \{R_1, R_2, , R_r\}$ such that for any region $R_j$, each query in $W$ selects either all records in $R_j$ or none. These regions are the fundamental regions of $R$ induced by $W$. STRAT tries to minimize the expected relative error of the queries, for which it makes stronger assumptions about the future queries. Specifically, it assumes that fundamental regions of future queries are identical to the FRs of past queries, where FR of a query is the exact set of tuples accessed by that query. For streaming data processing, since the data stream is continuous, random and varied, such strong assumptions does not hold.

SciBORQ [69] is a data-analytics framework designed for scientific workloads and provides precise control over runtime and quality of query answering. It uses uses special structures, called impressions. An impression is selected such that the statistical error of a query answer remains low, while the result can be computed within strict time bounds. Impressions differ from other sampling approaches in their bias towards the focal point of the scientific data exploration, their multi-layer design, and their adaptiveness to shifting query workloads. It provides a framework for scientific data exploration and discovery, capable of producing quality answers with strict error bounds in pre-defined time frames. SciBORQ does not provide any guarantees on the error margin.

ApproxHadoop [43] is another sampling based approximation system and uses random sampling technique for approximation. It targets persistent data processing and uses multi-stage sampling [57] for approximate MapReduce [34] job execution. It proposes various approximation mechanisms for MapReduce paradigm, such as input data sampling, task dropping, and running a precise, user-defined approximate version of the MapReduce code. ApproxHadoop also estimates error of distributed MapReduce programs that approximates with input data sampling or task dropping. Since it uses random sampling technique, the minority group might be neglected.

BlinkDB [12, 13] is another system that explores sampling based approximation and error estimation. It proposed an approximate distributed query engine that uses stratified sampling [14] to provides support for ad-hoc queries with error and response time constraints. It sufficiently represents the rare-subgroups using stratified sampling technique. BlinkDB is also suitable to ad-hoc workloads. It provides *(i)* an adaptive optimization framework that builds and maintains a set of multi-dimensional stratified samples from original data over time, and *(ii)* a dynamic sample selection strategy that selects an appropriately sized sample based on a query's accuracy or response time requirements.

All these aforementioned systems target approximate computing for batch processing. In contrast, our system targets low-latency stream processing. Furthermore, we extend the approximation with incremental computation.

# 8  Conclusion

In this thesis, we presented the marriage of incremental and approximate computations. Our approach transparently benefits unmodified applications, i.e., programmers do not have to design and implement application-specific dynamic algorithms or sampling techniques. We build on the observation that both computing paradigms rely on computing over a subset of data items instead of computing over the entire dataset. We marry these two paradigms by designing a sampling algorithm that biases the sample selection to the memoized data items from previous runs. We implemented our algorithm in a data analytics system called INCAPPROX based on Apache Spark Streaming. Our evaluation shows that INCAPPROX achieves improved benefits of low-latency execution and efficient utilization of resources.